# High efficiency near diffraction-limited mid-infrared flat lenses based on metasurface reflectarrays

Shuyan Zhang,[1,#] Myoung-Hwan Kim,[2,3,#] Francesco Aieta,[1,4] Alan She,[1] Tobias Mansuripur,[5] Ilan Gabay,[1] Mohammadreza Khorasaninejad,[1] David Rousso,[1,6] Xiaojun Wang,[7] Mariano Troccoli,[7] Nanfang Yu,[2] and Federico Capasso[1,*]

[1]John A. Paulson School of Engineering and Applied Sciences, Harvard University, 9 Oxford Street, Cambridge, MA 02138, USA.

[2]Department of Applied Physics & Applied Mathematics, Columbia University, 500 West 120th St, New York, NY 10027, USA.

[3]Department of Physics, The University of Texas Rio Grande Valley, Brownsville, TX 78520, USA.

[4]LEIA 3D, Menlo Park CA 94025, USA.

[5]Department of Physics, Harvard University, 17 Oxford Street, Cambridge, MA 02138, USA.

[6]University of Waterloo, Waterloo, ON N2L 3G1, Canada.

[7]AdTech Optics, Inc., 18007 Courtney Court, City of Industry, CA 91748, USA.

[#]These authors contributed equally.

*Corresponding author: capasso@seas.harvard.edu.

**A limiting factor in the development of mid-infrared optics is the lack of abundant materials that are transparent, low cost, lightweight, and easy to machine. In this paper, we demonstrate a metasurface device that circumvents these limitations. A flat lens based on antenna reflectarrays was designed to achieve near diffraction-limited focusing with a high efficiency (experiment: 80%; simulation: 83%) at 45° incidence angle at λ = 4.6 µm. This geometry considerably simplifies the experimental arrangement compared to the common geometry of normal incidence which requires beam splitters. Simulations show that the effect of comatic aberrations is small compared to parabolic mirrors. The use of single-step photolithography allows large scale fabrication.**

## 1. INTRODUCTION

The interest in the applications of mid-infrared (IR) light has recently increased in many areas, such as night vision, pharmaceutical quality monitoring, and homeland security. Over the last decade, there has been major progress in the development of new IR sources and detectors. However, a limiting factor in the development of mid-wave and long-wave IR optics is the lack of suitable materials that are transparent, low cost, lightweight, and easy to machine.

Recently, a new class of optical components based on metasurfaces has been developed [1]. These components control the wavefront of light using arrays of optical resonators with subwavelength dimensions, which are patterned on a surface to introduce a desired spatial distribution of optical phase. By tailoring the properties of each element of the array, one can spatially control the phase of the scattered light and consequently mold the wavefront. Based on this concept, various functionalities have been demonstrated including lenses [2, 3], axicons [2], blazed gratings [4], vortex plates [5] and wave plates [6]. In particular, high-efficiency flat lenses have been recently demonstrated in the transmission mode based on high-contrast gratings using hydrogenated amorphous silicon in the near-IR and mid-IR spectral range [7, 8].

In the reflection mode, designs based on reflectarray metasurfaces have been proposed to improve the efficiency and to realize different functionalities. A reflectarray metasurface is a multi-layered structure consisting of a planar array of antennas separated from a ground metallic plane by a dielectric spacer of subwavelength thickness. For example, reflectarrays with H-shaped antennas were designed for converting propagating microwaves to surface waves and patch antennas were designed to demonstrate anomalous reflection in the near-IR [9-11]. Reflectarray metasurfaces have also been designed for focusing in the visible [12] and the near-IR [3, 13], wide-angle holography in the near-IR [14], and polarization-dependent beam steering in the mid-IR [15] and the terahertz spectral range [16].

In this paper, we report the first mid-IR (λ = 4.6 µm) flat lens based on reflectarrays. They have high focusing efficiency (80%), the highest reported so far for reflectarray-based lenses. Experiments and simulations show that the focusing is near diffraction-limited. Importantly, the comatic aberration is small compared to its parabolic mirror counterpart. In addition, the lens is extremely thin and lightweight compared to conventional bulky optical components. Its material and processing requirements are minimal, requiring only a flat reflective substrate and one photolithography step using a high throughput deep-UV photolithography stepper, which enables large-scale fabrication.

## 2. LENS DESIGN

As a proof of principle, we demonstrated a flat lens with the functionality of a cylindrical lens (one-dimensional focusing). A schematic of the flat lens is shown in Fig. 1 (a): a collimated Gaussian beam arrives at the flat lens at an oblique incidence angle θ, and is focused at a length $f$ in the direction normal to the lens surface. The reason for the oblique incidence is purely practical: by using the metasurface itself to spatially separate the incident and reflected beams, we avoid the need for a beam splitter to separate the two beams. This technique simplifies the experimental setup and eliminates the insertion loss introduced by a beam-splitter [3, 13], which improves weak signal detection.

In order to achieve the desired focusing for θ = 0°, the phase profile of the wavefront as a function of position $x$ along the metasurface lens must satisfy [13]:

$$\boldsymbol{\varphi_{focus}(x)} = \frac{2\pi}{\lambda_0} \cdot \left(f - \sqrt{x^2 + f^2}\right), \quad (1)$$

where $\lambda_0$ is the wavelength. There is no phase modulation along the y-direction, resulting in a focal line rather than a point. The oblique incident beam arrives with a linear phase gradient of its own, which we can cancel out using the metasurface by applying an additional phase profile:

$$\boldsymbol{\varphi_{linear}(x, \theta)} = -\frac{2\pi}{\lambda_0} \cdot [x \cdot \sin(\theta)]. \quad (2)$$

Thus, we engineer the total phase shift $\varphi$ of the lens to be

$$\boldsymbol{\varphi(x) = \varphi_{focus}(x) + \varphi_{linear}(x)}. \quad (3)$$

The total phase function was realized by subwavelength antennas of fixed center-to-center separation. The unit cell of the antenna array is shown in Fig. 1 (b). We take advantage of the interaction of the antenna with its mirror image in the ground plane: the near-field coupling between the antenna and its image results in a reflected field with a broad phase coverage without polarization conversion [10]. At each center position $x$, the disc radius was chosen such that its phase response was closest to the calculated value $\varphi(x)$ (Supplementary Fig. S1).

Using this method, we designed a flat lens for the parameters: θ = 45°, f = 8 cm, and W = 3.08 mm in both x- and y-directions. The numerical aperture (NA) of our lens is 0.02, which can be increased by decreasing the focal length or increasing the size of the lens itself [2, 3]. We fabricated the flat lens on a 6-inch fused silica wafer substrate. The deposition of the gold (Au) back reflector layer and discs was done by electron-beam evaporation. The silicon dioxide ($SiO_2$) dielectric spacer layer was deposited using plasma-enhanced chemical vapor deposition. The antenna patterns were fabricated using a stepper photolithography tool (GCA AS200 i-line stepper), which allows for large-scale fabrication. Fig. 1 (c) shows a scanning electron microscope image of a small section of the fabricated lens. Note that the 6-inch wafer substrate can accommodate over 160 such lenses with different focusing properties, such as focal lengths and reflection angles.

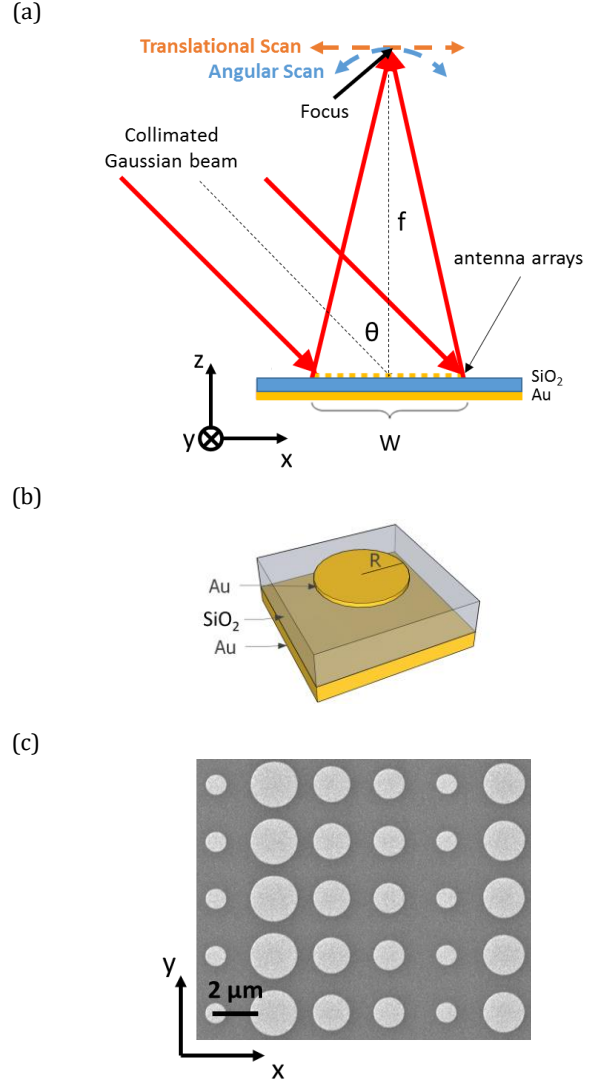

Figure 1: (a) Geometry of the flat lens based on reflectarray antennas. W is the width of the lens. The focus is a narrow line in the y-direction which is characterized by angular and translational scans indicated by the blue and orange dashed lines, respectively. The drawing is not to scale. (b) Schematic of the unit cell of the reflectarray lens: a 50 nm-thick gold disc is separated from a 200 nm-thick gold back reflector layer by a 400 nm-thick $SiO_2$ spacer. The size of the unit cell is 2.5 μm x 2.5 μm. (c) Scanning electron microscope image of a small section of the fabricated antenna arrays. The center-to-center distance of the discs is the same.

## 3. RESULTS AND DISCUSSIONS

We have performed both numerical simulations (FDTD module, Lumerical Inc.) and experimental measurements to characterize the performance of the flat lens.

Simulation results are shown in Fig. 2. Since there is no spatial variation of the antenna array in the y-direction, only one row of disc antennas was simulated and Bloch boundary conditions were applied in the y-direction. To monitor the phase profile created by the lens with an oblique incidence, a monitor was placed $2\lambda_0$ away from the lens surface to record data for the propagating waves only. The result is shown in Fig. 2 (a). The fuzziness is mainly due to the incomplete phase coverage of antennas (Supplementary Fig. S1) however the effect is small. The correlation coefficient of the curve fit is $R^2 = 0.99$, indicating a very good agreement between the designed phase profile and the ideal phase profile (black solid line) calculated from the design equation (1). Discussions of the simulation data and the related phase discretization error are presented in Supplementary Fig. S2. Fig. 2 (b) is the calculated distribution of the electric field intensity (normalized $|E|^2$) near the focal region in the x-z plane. It was obtained by propagating the field data obtained from the near-field monitor to the far-field (see [17]). Fig. 2 (c) shows the reflected beam intensity (normalized $|E|^2$ in log scale) from the flat lens along a semicircle of radius 8 cm, equal to the focal length of the lens. Only a focused beam at $\theta = 0°$ is observed. We have verified this experimentally (see Supplementary Fig. S3).

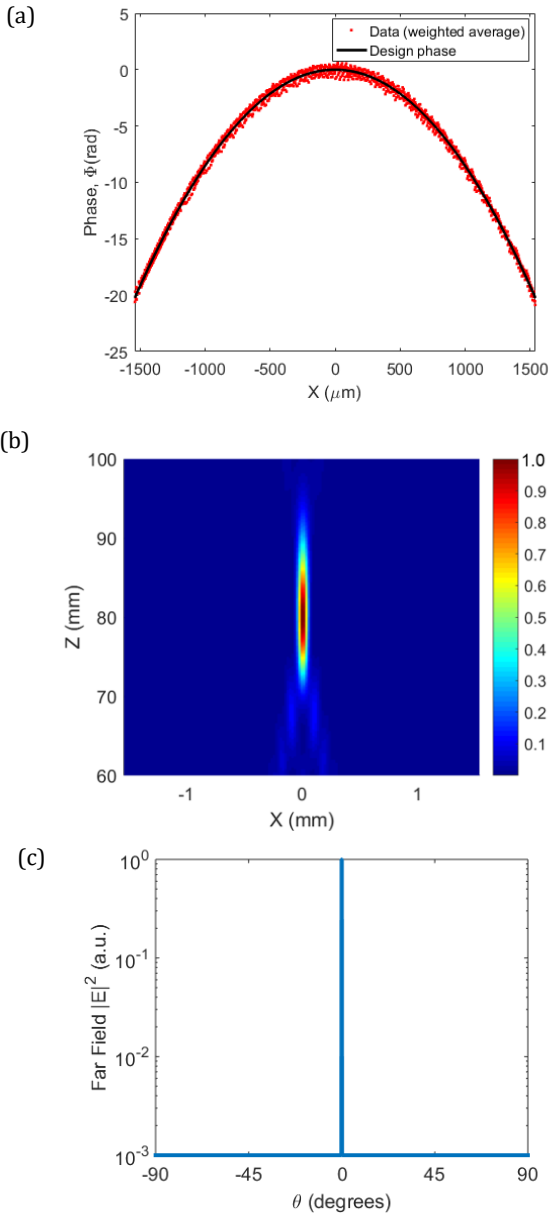

(a)
(b)
(c)

Figure 2: FDTD simulation results of the flat lens. (a) Phase profile created by the lens for an oblique incidence at $\theta = 45°$. (b) Distribution of the intensity (normalized $|E|^2$) of the reflected beam in the x-z plane at y = 0. The lens is centered at x = 0 and the size of the lens is from -1.54 mm to 1.54 mm. (c) The normalized far-field intensity of the reflected beam.

For the experimental characterization of the lens, we used a continuous-wave Fabry-Pérot quantum cascade laser (QCL) (AdTech Optics) emitting at $\lambda = 4.6$ μm. The laser was mounted so that its output beam is s-polarized (electric field along y-direction). We operated the laser near threshold so that the emitted beam is at a single mode. The full beam waist of the laser was 3 mm. The focusing beam created by the reflectarray was collected by a thermoelectrically-cooled mercury-cadmium-telluride detector (VIGO Systems) mounted on motorized rotation and translation stages. To increase the spatial resolution of the scans the light was first sent through a 30 μm pinhole before reaching the detector. The signal-to-noise ratio was increased by modulating the intensity of the QCL with a small sinusoidal current superimposed on the dc current (Wavelength Electronics QCL1500) and demodulating the detected signal with a lock-in amplifier (AMETEK Advanced Measurement Technology).

The reflected focused light from the flat lens was characterized experimentally by angular and translational scans as shown in Fig. 3 (a) and Fig. 3 (b), respectively. The data were taken at the center of the focal line, i.e. y = 0. The experimental results are in good agreement with the simulations. The average standard deviation of repeated measurements is 0.2%. Note that the step sizes used in scanning the reflected beam in Fig. 3 (a) was 0.01° (equivalent to 14 μm in the x-direction) and in Fig. 3 (b) was 10 μm, both smaller than the pinhole diameter (30 μm), so the raw data are a convolution of the true signal and the pinhole response function. Hence, deconvolution was performed to retrieve the original beam profiles. Fig. 3 (a) shows that the measured reflection angle is 0°, which agreed well with the design equation. The full width at half maximum (FWHM) is less than 0.1°. Fig. 3 (b) shows that the full beam waist at $1/e^2$ of the peak intensity (indicated by the two black arrows) is 164 μm in the experiment and 166.5 μm in the simulation. The diffraction-limited full beam waist ($2w_0$) is 156 μm, calculated by Fourier transformation of an ideal continuous focusing phase profile. Our measured focused beam size is close to the diffraction limit. The difference is likely due to the incomplete phase coverage and the variation of the reflectance of the antennas (Supplementary Fig. S1). Since our lens focuses light into a line, we checked the focused beam profile at other y positions and verified that the focus is near diffraction-limited along the focal line (see Supplementary Fig. S4). It is worth pointing out that although the results shown are for s-polarized incident light, we have simulated and measured the lens with p-polarized incident light and found that $2w_0 = 167$ μm. Hence, the focusing performance of our lens is near polarization-independent (Supplementary Fig. S5).

The measured focusing efficiency of the lens was $\eta = 80\%$ which is close to the simulated value of 83%. Most of the loss comes from the absorption of the materials. Experimentally, the focused beam power was measured by placing a laser power meter (Nova, Ophir Photonics) in the focal region with an iris in front to block non-focused light. Note that since the incidence angle was 45°, the effective incident beam size along the lens surface was bigger than the size of the lens, hence, the portion of the incident power on the lens was 82%, which was used to calculate the focusing efficiency. In the simulation, $\eta$ was calculated by taking the ratio of the optical power in the focal region in Fig. 2 (c) and the incident power. The discrepancy is likely due to fabrication errors.

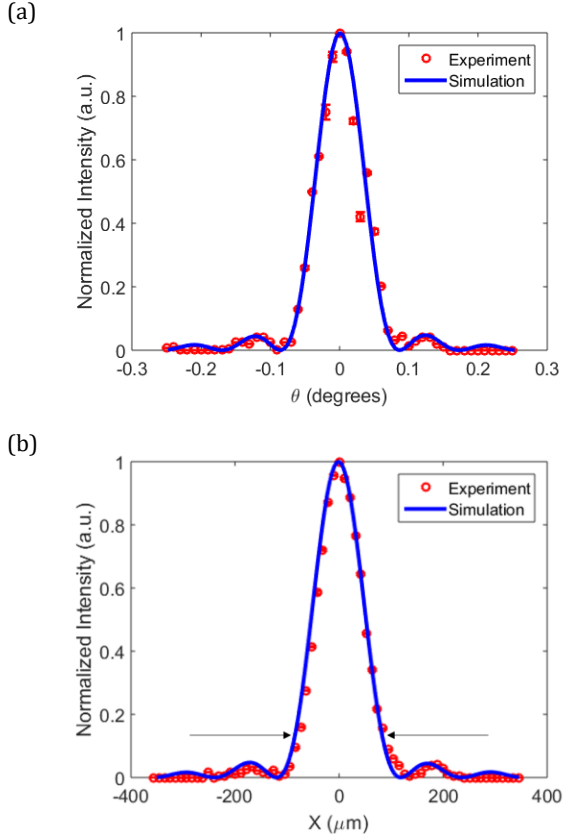

Figure 3: Angular scan (a) and translational scan (b) of the reflected beam intensity (normalized $|E|^2$) measured at the center of the focal line. The arrows in (b) indicate the full beam waist measurement in μm.

## 4. EFFECTS OF INCIDENCE ANGLE AND WAVELENGTH

This section discusses the effect of comatic and chromatic aberration of our flat lens. Detailed discussions on aberrations of flat lenses can be found in [18].

Table 1 shows the comatic aberration of the lens. The wavelength is fixed at 4.6 μm. The sign of the angle is defined in Fig. 1 (a). If the incidence angle increases (or decreases) from the design angle ($45°$), the reflected light shifts to more negative (or positive) angles from the normal to the lens surface. The comatic aberration has the following physical origin: any deviation from the designed incidence angle introduces an extra phase gradient on the surface of the lens, which cannot be compensated by the phase profiles of the antennas. Therefore, scattered waves from each antenna will not arrive at the designed focal line in phase, resulting in a shift in the position of the focal line. The focal length and the full beam waist, however, remain relatively constant as the incidence angle changes within ±15° from the design angle.

The beam profiles at the center of the focal line for various incidence angles in Table 1 obtained by FDTD simulations are plotted in Fig. 4 (a). The inset shows the difference between each curve taking the one at θ = 45° as the baseline. The difference between the beam profiles is less than 8%. We have cross-checked the results with experiment for θ = 35° and 55°, which are shown in Fig. 4 (b). The simulations and experiments agree well with each other.

Table 1. Comatic aberrations at $\lambda_0 = 4.6$ μm.

| Incidence angle (°) | Reflection angle (°) | Focal length (cm) | Full beam waist $1/e^2$ (μm) |
|---|---|---|---|
| 30 | 12 | 7.6 | 163 |
| 35 | 7.7 | 7.8 | 165 |
| 40 | 3.7 | 8 | 167 |
| 45 | 0 | 8 | 166.5 |
| 50 | -3.4 | 7.9 | 165.5 |
| 55 | -6.5 | 7.9 | 165.5 |
| 60 | -9.2 | 7.8 | 164.5 |

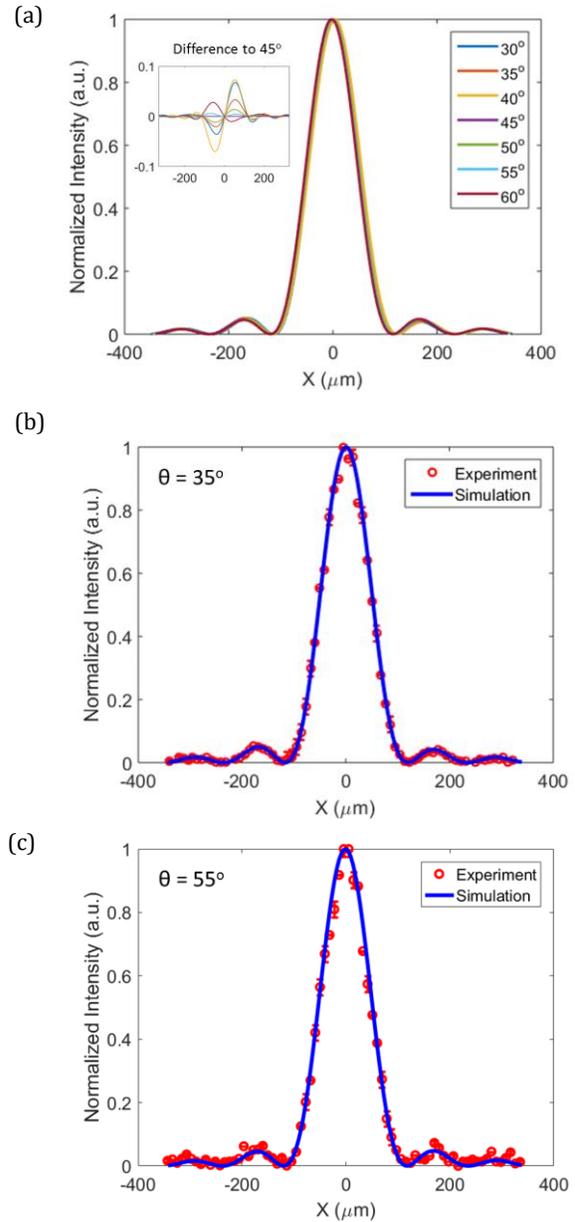

Figure 4: (a) The focused beam profile for various incidence angles obtained by FDTD simulations. The inset shows the difference between the beam profiles and the one corresponding to an incidence

angle of θ = 45°. Experimentally measured focused beam profiles for (b) θ = 35° and (c) θ = 55° compared with the simulation results.

Parabolic mirrors are widely used for focusing in the IR spectral range but are known to be very sensitive to comatic aberrations. Hence, we compared the comatic aberrations of an off-axis parabolic mirror and our lens with the same NA using ray tracing techniques. The result is shown in Table 2. With 10° change in the incidence angle, the change in the reflection angle of the parabolic mirror is -10° as expected, but that for the reflectarray lens is smaller, only -6.5°. The percentage change in the focal length is also smaller for the reflectarray lens, 1%, compared to the parabolic mirror which is 9%. In summary, our reflectarray lens is not only thinner and lighter, but also can significantly reduce the comatic aberration as compared to a parabolic mirror.

**Table 2. Comparison of reflectarray lens and parabolic mirror with NA = 0.02.**

| Type | Change in incidence angle (°) | Change in reflection angle (°) | Change in focal length |
|---|---|---|---|
| Reflectarray lens | 10 | -6.5 | 1% |
| Parabolic mirror | 10 | -10 | 9% |

Studies of the chromatic aberrations of the flat lens using FDTD simulations are summarized in Table 3. Depending on whether the incident wavelength is shorter or longer than $\lambda_0$ = 4.6 μm; the reflection angle is negative or positive, respectively. The focal length decreases but the full beam waist remains relatively constant as the incident wavelength increases.

**Table 3. Chromatic aberrations for incidence angle θ = 45°.**

| λ (μm) | Reflection angle (°) | Focal length (cm) | Full beam waist $1/e^2$ (μm) |
|---|---|---|---|
| 4.0 | -5.3 | 9.1 | 165 |
| 4.6 | 0 | 8 | 166.5 |
| 5.0 | 3.6 | 7.3 | 167 |

## 5. CONCLUSIONS

In this paper, we designed a mid-IR flat lens that consists of a reflectarray of metallic antennas with subwavelength spacing. The focusing performance and aberrations of the lens were studied by experiments and simulations. We found that the effects of comatic aberrations on the focal length and focused beam size were small for a deviation of the incidence angle within ±15°, which compare favorably to an off-axis parabolic mirror with the same NA. Our approach based on illuminating the flat lens at an oblique angle and achieving near diffraction-limited focusing at the lens surface normal direction has the advantage of a simplified experimental setup compared to other reflection-based measurements with normal incidence. Our lens has a number of attractive features, including high focusing efficiency, flexibility of design, polarization-independence, straightforward fabrication based on single-step photolithography, and reproducibility on a large scale.

Although our flat lens is not aberration-free, there are strategies to reduce comatic [18, 19] and chromatic aberrations [20-25] of flat lenses. We envision that our flat lenses will complement or replace various conventional optical components in systems for IR imaging, ranging, and detection, as well as for beam shaping of IR lasers and beacons.

**Funding.** Air Force Office of Scientific Research (AFOSR) (MURI: FA9550-12-1-0389), National Science Foundation (ECCS-1307948), and Draper Laboratory (SC001-0000000731). Shuyan Zhang acknowledges the funding support from National Science Scholarship from A*STAR, Singapore.

**Acknowledgment**. We acknowledge Mikhail A. Kats and Steven Byrnes for helpful discussions and Elodie Strupiechonski for the valuable advice on the experiments. This work was performed in part at the Center for Nanoscale Systems (CNS), a member of the National Nanotechnology Infrastructure Network (NNIN), which is supported by the National Science Foundation under NSF award no. ECS-0335765. CNS is part of Harvard University.

See Supplement 1 for supporting content.

# Supplementary Information

### 1. Phase response of antennas

We used FDTD simulations with periodic boundary conditions to find a set of phase shift elements with different radii that yielded both a large range of reflection phase and a relatively uniform reflection amplitude, as shown in Fig. S1. The unit cell configuration is shown in Fig. 1 (b) in the main manuscript. The phase coverage is from 0 to approximately 1.6π for discs with radii ranging from 0.4 μm to 1 μm. We note that the phase and reflectance are weakly dependent on the incidence angle, so we have tested two lens designs based on the phase response at the incidence angles of 0° and 45° and we found that the focusing performance is almost the same.

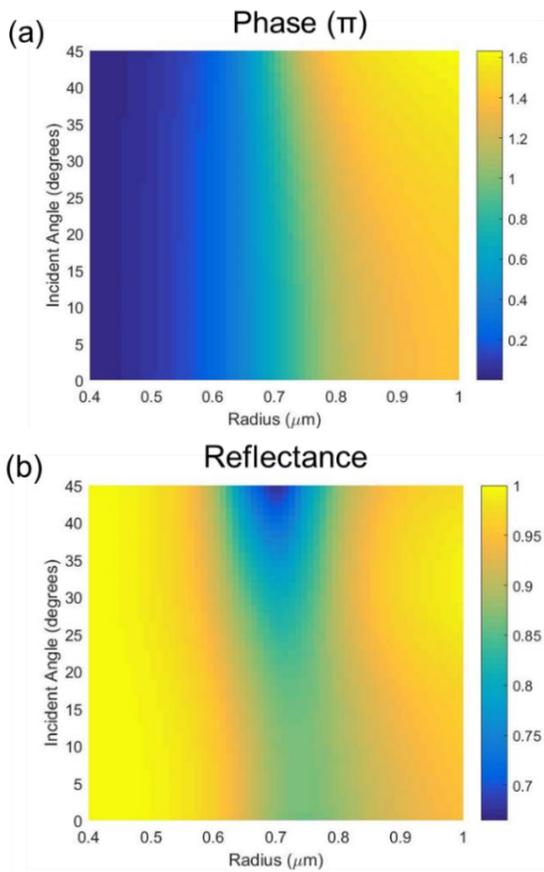

Figure S1: (a) Phase and (b) amplitude response of antennas with varying disc radius for different incidence angles from 0° to 45°.

### 2. Phase discretization error

The realization of the phase profile calculated from the equations is through discrete antennas, hence each antenna has to cover the phase for a particular area, i.e. the size of the unit cell. Fig. S2 is the simulation data (blue and orange dots) of the phase and amplitude of the electric field of a portion of the lens with incident light. The spatial resolution of the monitor was 0.1 μm. The data were obtained from a monitor $1/5\lambda_0$ away from the lens top surface. The reason why the monitor was placed at a near-field distance is that after a certain propagation distance, the phase at each position will be a combination of the phase from neighboring antennas, but we want to study the phase discretization error caused by each individual antenna itself. The x-axis is from -800 μm to -750 μm covering 20 unit cells. The size of the unit cell (2.5 μm) is marked by the black dashed lines. The blue circles represent the sampling of the ideal phase profile given by equation (1) at each antenna position. The variation of the simulated phase profile within each unit cell as compared to the ideal phase profile leads to the phase discretization error. The closer the antennas are, the smaller the error is, but the size of the unit cell is limited by factors such as inter-antenna coupling and fabrication resolution. Because of the phase and amplitude variation within the unit cell, Fig. 2 (a) in the main manuscript shows the simulation data weighted by the amplitude of the electric field at each point and averaged over a 2.5 μm interval spacing.

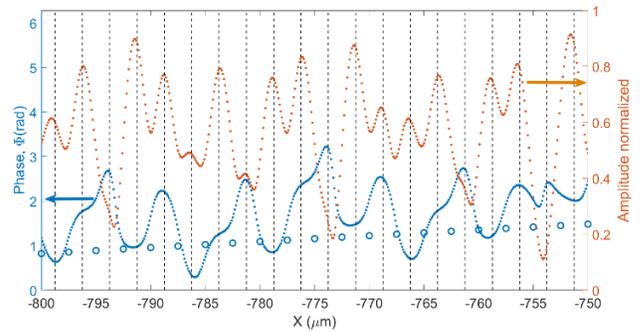

Figure S2: Simulated near-field phase and amplitude distribution (plotted is a portion of the lens). The blue and orange dots are data points of the phase and amplitude, respectively. The blue circles are the sampling of the design phase profile by the antenna positions. The black dashed lines mark the size of the unit cell.

### 3. Reflected beams from the flat lens

Fig. S3 shows the experimental results of an angular scan ranging from -50° to 10° with a step size of 1°. The range of the angular scan is limited by the experimental setup. In addition to the beam focused by the flat lens at θ = 0°, we also observed a reflected beam at θ = -45°. This is because the effective laser beam size along the lens surface was bigger than the lens; as a result, the portion of the light not incident on the lens was directly reflected by the gold backplane. The intensity of the light at -45° is approximately 16.2% of the laser output power.

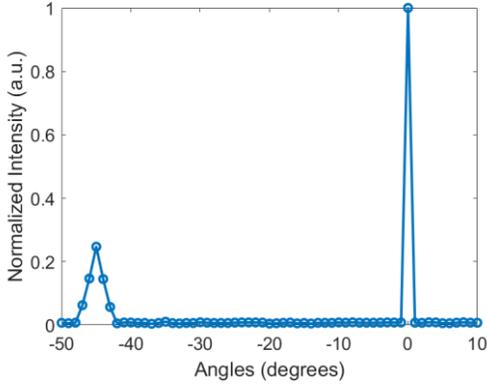

Figure S3: Angular scan of the reflected beam intensity (normalized) from -50° to 10° at y = 0.

## 4. Characterization of the lens performance along the focal line

Since our lens is a cylindrical lens, the focus is a focal line. The data presented in the main manuscript were taken at y = 0. We expect that the focused beam profile does not change along the y-direction. We experimentally verified it at two other y positions, which are shown in Fig. S4 (a) and (b). The full beam waist size is 166 μm and 165 μm at y = -0.75 mm and y = 1.4 mm, respectively. Recall that the full beam waist at y = 0 is 164 μm. The difference is small, so we conclude that there is no aberration along the focal line. However, we note that the focused beam intensity varies along the focal line as shown in Fig. S4 (c). This is due to the intensity variation of the Gaussian beam source. The intensity drops to nearly zero beyond the size of the lens, i.e. -1.54 mm to 1.54 mm.

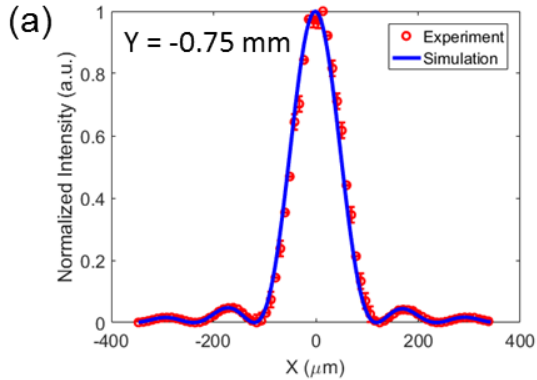

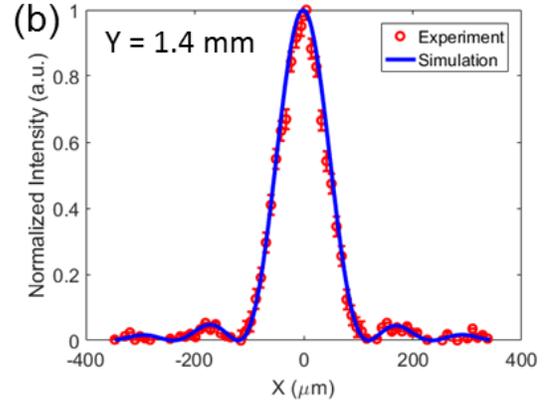

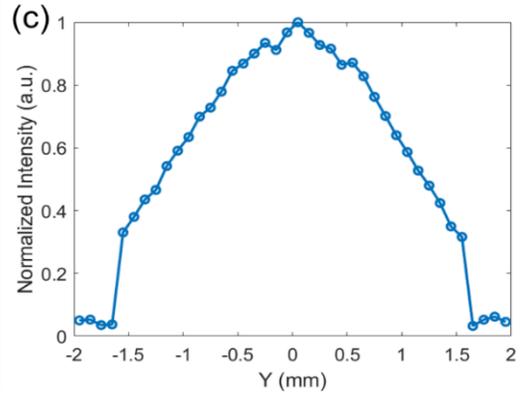

Figure S4: Normalized focused beam profile at (a) y = -0.75 mm and (b) y = 1.4 mm. (c) Normalized focused beam intensity as a function of y measured at x = 0. The step size of the measurement is 100 μm.

## 5. Focusing performance of p-polarized light

This section summarizes the simulation and measurement results when the incident light is p-polarized (electric field in the plane of incidence) with an incidence angle of θ = 45°. Experimentally, this was realized by mounting the QCL vertically. All measurements were performed on the same lens as in the main manuscript. We found that the focusing performance including the focal length, the beam waist at the focus, and the focusing efficiency is similar to that of the s-polarized light.

Fig. S5 shows the FDTD simulations of the flat lens with p-polarized illumination. Fig. S5 (a) is the distribution of the electric field intensity (normalized $|E|^2$) around the focal region in the x-z plane. The focal length is 8 cm. Fig. S5 (b) shows the reflected beam intensity (normalized $|E|^2$ in log scale) along a semicircle of radius of 8 cm. The focusing efficiency is 82% (simulation) and 78% (experiment), which is slightly lower than that of the s-polarized light.

The reflected focused light from the flat lens was characterized experimentally by angular and translational scans as shown in Fig. S5 (c) and Fig. S5 (d), respectively. The experimental results are in good agreement with the simulations. The scanning parameters are the same as that for the s-polarized light and deconvolution was performed. The reflected light was focused at 0° and the full beam waist ($2w_0$) at the focus was found to be 167 μm, both in the experiment and in the simulation.

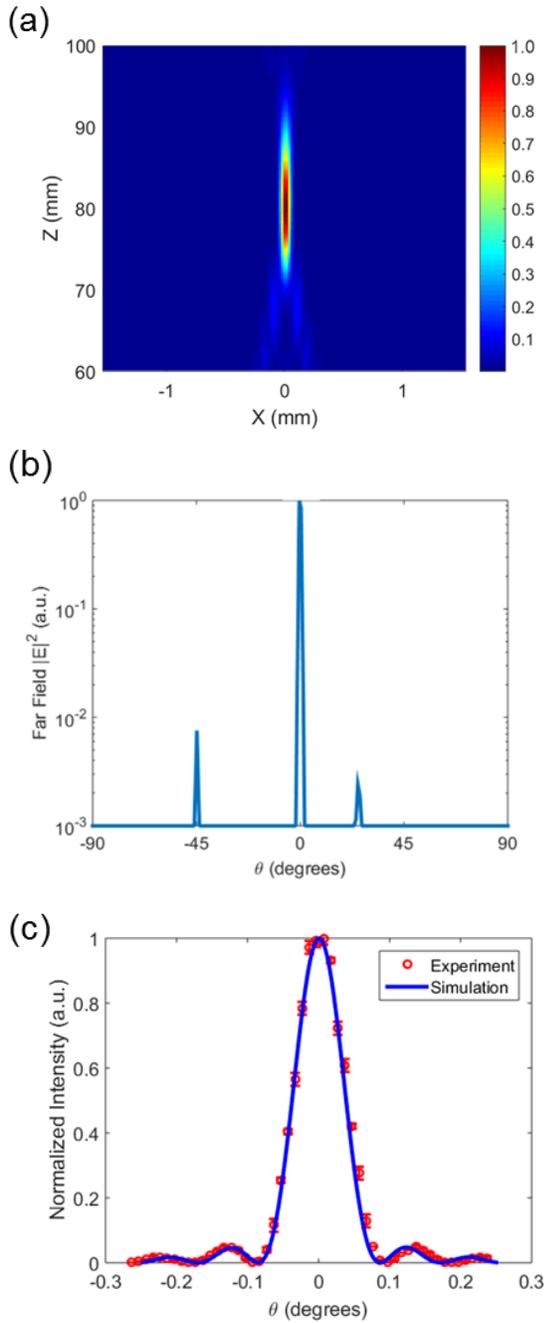

Figure S5: Simulation and experimental results of the flat lens with p-polarized incident light. (a) Distribution of the intensity (normalized $|E|^2$) of the reflected beam in the x-z plane. The antenna arrays are centered at x = 0. (b) Far-field projection of the reflected beam intensity (normalized $|E|^2$ in log scale) showing that the majority of the light was focused at 0° (normal to the lens surface). (c) Angular scan and (d) Translational scan of the reflected beam intensity (normalized $|E|^2$) at the focus.